\documentclass[twocolumn,apj, pra, showpacs,preprintnumbers,amsmath,amssymb]{revtex4}

\usepackage[pdftex]{graphicx} 
\usepackage{epsfig}
\usepackage{dcolumn}
\usepackage{bm}
\usepackage{wrapfig}
\usepackage{times}

\begin{document}

\title{Multifrequency spin resonance in diamond}

\author{Lilian Childress}
\author{Jean McIntyre}
\affiliation{Department of Physics and Astronomy, Bates College, Lewiston ME}%

\date{\today}

\begin{abstract}
Magnetic resonance techniques provide a powerful tool for controlling spin systems, with applications ranging from quantum information processing to medical imaging.  Nevertheless, the behavior of a spin system under strong excitation remains a rich dynamical problem.  In this paper, we examine spin resonance of the nitrogen-vacancy center in diamond under conditions outside the regime where the usual rotating wave approximation applies, focusing on effects of multifrequency excitation and excitation with orientation parallel to the spin quantization axis.  Strong-field phenomena such as multiphoton transitions and coherent destruction of tunneling are observed in the spectra and analyzed via  numerical and analytic  theory.  In addition to illustrating the response of a spin system to strong multifrequency excitation, these observations may inform techniques for manipulating electron-nuclear spin quantum registers. 

\end{abstract}

\maketitle
\section{Introduction}


   Control over individual electronic and nuclear spins in the solid state offers promise for applications in  metrology and quantum information science~\cite{Dutt07, Neumann08, Jiang09, Morton08, Jones09, Baugh06}.  Manipulation of such systems often utilizes a combination of microwave (MW) excitation to drive electronic spin transitions and radio-frequency (RF) excitation to control nuclear spins~\cite{Neumann08, Jiang09, Smeltzer09, Steiner10}.  Experimental techniques typically rely on sequential RF and MW pulses~\cite{Neumann08, Jiang09}, but a better understanding of simultaneous MW and RF excitation may enable greater parallelism in manipulating few-spin systems.  Moreover,  multifrequency excitation can lead to a variety complex behaviors of fundamental interest~\cite{Haroche70}.  At the same time, control techniques using strong-field resonance phenomena have attracted attention in systems ranging from trapped atoms~\cite{Lignier07} to superconducting devices~\cite{Son09} as well as single spins in the solid state~\cite{Fuchs09}.  As transition rates approach excitation frequencies, complex dynamics ensue, and theoretical techniques beyond the usual rotating wave approximation are required to interpret the system evolution~\cite{Ashhab07}.  In this paper we consider a combination of these effects by examining multifrequency excitation of a single spin system in a regime where strong-field phenomena occur.   


We approach this problem from the point of view of recent experiments on the nitrogen-vacancy (NV) center in diamond~\cite{Neumann08, Jiang09, Dutt07}.  In particular, we consider a situation directly relevant to the study of electron and nuclear spin resonance in diamond: simultaneous excitation by weak MW intended to drive electronic spin transitions and strong RF intended to drive nuclear spin resonance~\cite{Smeltzer09}.  Our experimental observations are dominated by the response of the electronic spin, and exhibit features characteristic of strong-field excitation such as multiphoton resonances and coherent destruction of tunnelling. 
By comparison to numerical simulations and analytic theory, we determine that the behavior we observe involves both transverse and longitudinal components of the magnetic fields driving the NV spin.  Beyond explaining the observed resonances, the identification of multiaxis coupling also suggests techniques for controlling the NV spin system that go beyond the usual transverse spin-resonance model.  

\section{Multifrequency excitation of the NV center in diamond}


The electronic spin associated with NV center in diamond has recently emerged as a promising candidate for quantum information processing~\cite{Neumann08, Jiang09, Dutt07} and  sensitive magnetometry~\cite{Taylor08, Maze08mag, Bala08}.  Its diamond host offers a stiff, nearly spinless environment, leading to extremely long electronic spin coherence times in the range of a millisecond~\cite{Bala09}.  Moreover, the optical transitions of the NV center allow a high degree of spin polarization at room temperature via optical pumping, and enable spin-dependent fluorescence for optical spin detection~\cite{Manson06}.  With several possible nuclear spins in its environment -- $^{14}$N or $^{15}$N and isotopic impurity $^{13}$Cs -- the NV center offers the opportunity to prepare, manipulate, and measure a single NMR molecule in the solid state.  

\begin{figure}[bhtp] 
  \centering
  \includegraphics[width=3.25 in,height=5in,keepaspectratio]{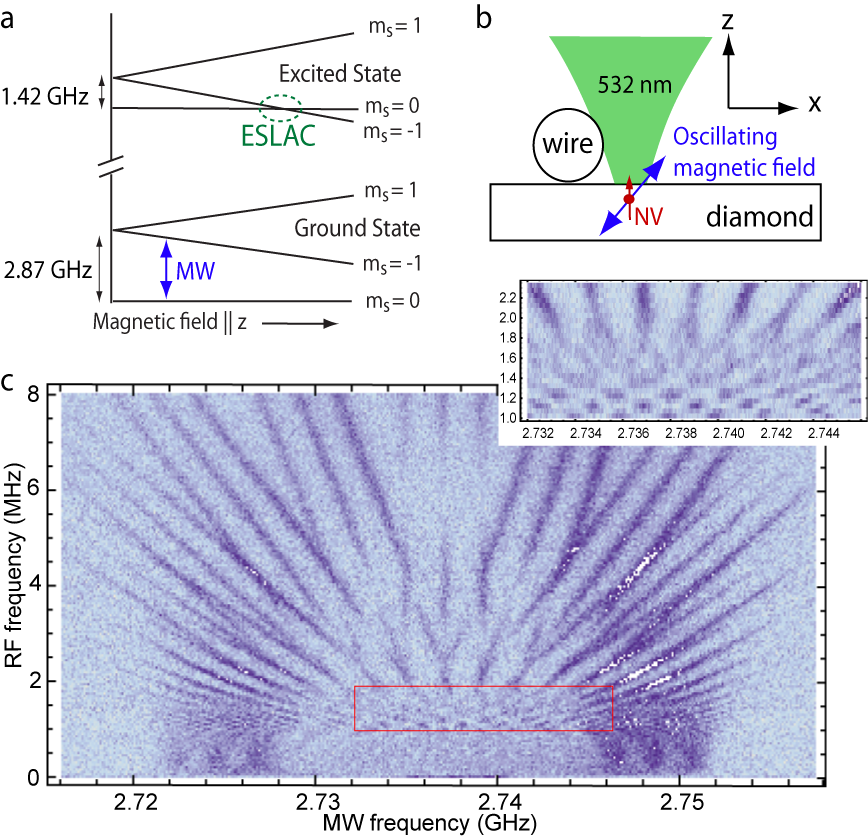}
  \label{geom}
\caption{(a) The spin levels of the electronic ground state and optically excited state of the NV center.  Note that the orbital doublet in the excited state is averaged at room temperature  where we conduct our experiments~\cite{Rogers09}.  The excited state level anticrossing (ESLAC) near 510G is indicated by a dashed oval. (b) The geometry of our experiments, illustrating the orientation of oscillating magnetic fields. (c) Multifrequency continuous-wave excitation in low magnetic field.  Inset shows zoom of boxed region.  By extrapolating from Rabi nutation frequencies at higher MW intensity, we estimate the MW electron spin Rabi frequency for this data to be $\Omega_{MW}= (2\pi)0.46$ MHz.  The RF intensity is significantly higher; based on quasistatic analysis (below), we estimate the RF longitudinal coupling to be $\Omega_{RF} \approx (2\pi) 6$ MHz.  Each horizontal sweep in MW frequency is normalized by its average value, compensating for slow fluctuations in fluorescence intensity owing to sample drift as the data is collected over several days.}
\end{figure}

The ground state of the NV center has an electronic spin $S = 1$ with a natural spin quantization axis $S_z$ provided by the substitutional nitrogen and adjacent vacancy that comprise the defect.  In the absence of a magnetic field, the ground state $m_s = 0$ spin sublevel lies 2.87 GHz below the $m_s = \pm 1$ levels; in the presence of a magnetic field, the NV spin gyromagnetic ratio is extremely close to that of the bare electron.  The optically excited state of the NV defect likewise has an $S = 1$ configuration with the same quantization axis and similar gyromagnetic ratio, but its room temperature zero-field splitting is only $1.42$ GHz~\cite{Fuchs08} (see Fig. 1a).  While electron spin resonance (ESR) has been observed in the excited state~\cite{Fuchs08, Steiner10}, the fast optical decay time $\tau \sim 12$ ns~\cite{Manson06} limits the resolution of such experiments, and the experiments discussed below involve spin resonance in the ground state configuration.

\subsection{Experimental observations}

In our experiments, single NV centers are isolated using a home-built confocal microscope operating at 532nm for excitation and 640-800nm for detection.  Helmholtz coils or a permanent magnet are used to apply static magnetic fields to the sample, while oscillatory fields are coupled onto the diamond by a $25\mu$m copper wire soldered across its surface.  
We image an NV center a few $\mu$m below the (111) surface of the diamond, choosing an NV center whose axis is perpendicular to the surface.  Currents through the 25$\mu$m copper wire create oscillating MW and RF magnetic fields oriented at an angle to the NV axis as illustrated in Fig~1b.  
 Furthermore, because of the proximity of the narrow wire to the NV center, the amplitude of the applied oscillatory magnetic fields can exceed 10 Gauss.   

We examine the NV center spectrum in a multifrequency parameter regime relevant for manipulation of coupled electronic and nuclear spin systems. 
Because hyperfine constants for nuclear spins in close proximity to the NV center are typically in the range of a few MHz (e.g. 2.2 MHz for $^{14}$N, 3 MHz for $^{15}$N~\cite{Felton09}), a combination of few-GHz and few-MHz excitation is required to resonantly drive both spins~\cite{Jiang09, Smeltzer09}.  Owing to tiny nuclear gyromagnetic ratios, the RF oscillatory magnetic field must be quite large (many Gauss) to enable nuclear spin transitions on $\mu$s timescales, whereas the MW fields driving electronic spin transitions may be 10 to 100 times weaker to selectively drive individual hyperfine lines.  We examine simultaneous multifrequency illumination in these frequency and power regimes, and find that it produces a complex set of resonances dominated by the electron spin response.

Fig.~1c shows a typical multifrequency spectrum for an NV center in diamond at low magnetic field, wherein an NV center in a magnetic field of 45 G $||\hat{z}$ is illuminated by 532nm light while MW and RF currents oscillate in the wire above it.  The fluorescence intensity (in the color scale, darker indicates lower fluorescence) is recorded as a function of RF and MW frequency ($\omega_{RF}$ and $\omega_{MW}$); darker regions indicate population driven from the $m_s = 0$ state into the $m_s = -1$ state.  In the absence of RF excitation, the electron spin resonance (ESR) spectrum at this magnetic field  would show three hyperfine lines from the three spin projections of the $I = 1$ $^{14}$N nuclear spin, centered around 2.737 GHz.  With RF excitation, three significant features appear:  At low RF frequencies, the spectrum is split by $24$ MHz; for $\omega_{RF} \geq (2 \pi) 5$ MHz, multiphoton phenomena appear; at moderate RF frequencies, the structure becomes complex (see inset) owing to overlap of hyperfine lines, multiphoton transitions, and ``missing" resonances such as the one that occurs near $\omega_{RF}= (2\pi) 5.5$ MHz, $\omega_{MW} = (2\pi) ~2.737$GHz.   

\begin{figure}[bhtp] 
  \centering
  \includegraphics[width=3.5 in,height=6in,keepaspectratio]{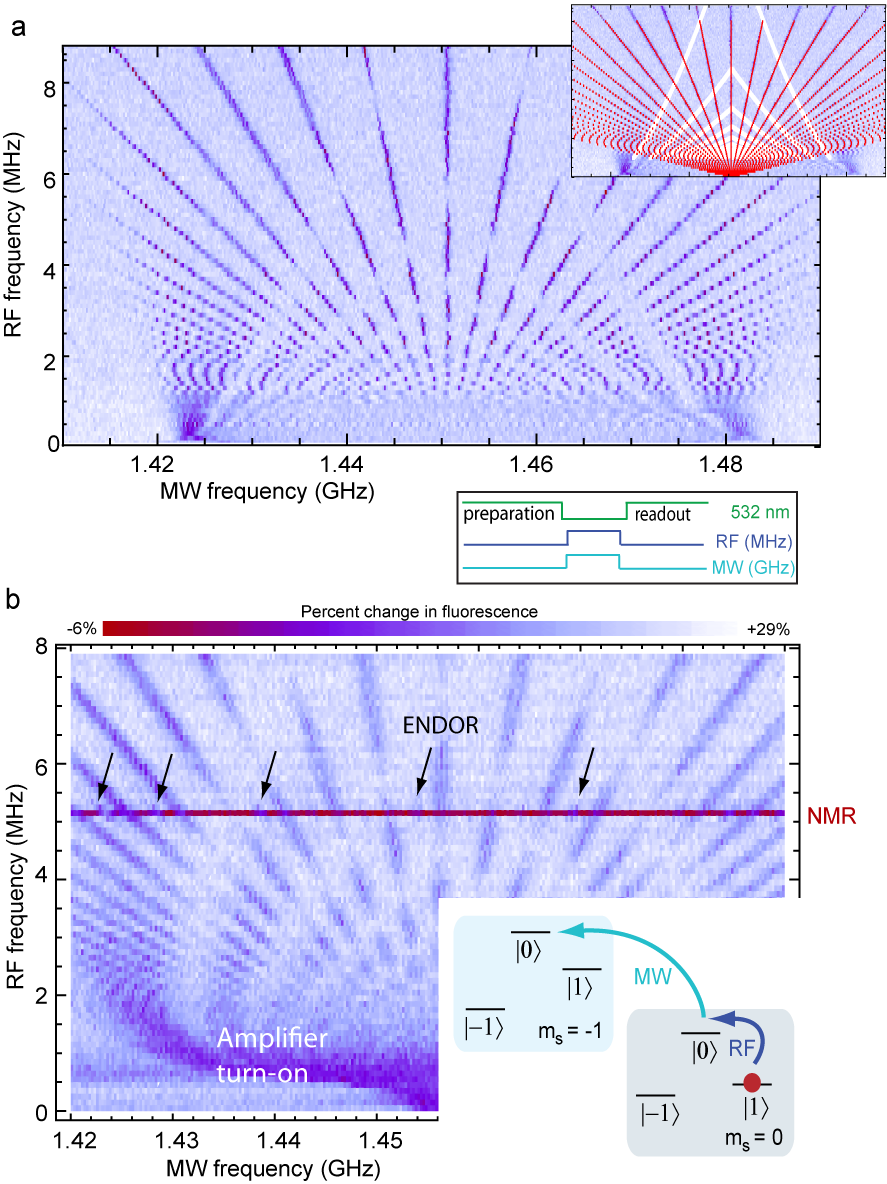}
  \label{peacock}
\caption{  (a) Multifrequency spectrum at the ESLAC. $\Omega_{RF} \approx (2\pi) 15$ MHz and $\Omega_{MW} \approx (2\pi) 0.43
$ MHz are pulsed for 5 $\mu$s prior to fluorescence detection (sequence shown).  Each horizontal sweep in MW frequency is normalized by its average value and shifted in frequency (using ESR data collected at regular intervals during data acquisition) to compensate for drifts in fluorescence intensity and magnetic field.  Inset shows data superposed with the lines $\delta = n\omega_{RF}$ (red) and zeroes of $J_{\delta/\omega_{RF}}\left(2\Omega_{RF}/\omega_{RF}\right)$ (white) for $\delta= \omega_{MW}-(2\pi)1.4506$ GHz and $\Omega_{RF} = (2\pi) 15.4$ MHz. (b) NMR and ENDOR at the ESLAC.  Higher intensities $\Omega_{MW} \approx (2\pi) 2$ MHz, $\Omega_{RF} \approx (2\pi) 20 $MHz and longer pulses (10$\mu$s) allow NMR and ENDOR via the mechanism illustrated in the inset.  Examples of weak ENDOR resonances are marked with arrows.  Note that the ENI 525LA RF amplifier used in this experiment does not have a flat response below 2 MHz.   
}
\end{figure}

The multifrequency spectrum of Fig.~1c can be simplified by polarizing the $^{14}$N nuclear spin.  Fig.~2 shows data taken at a magnetic field of $\sim 510$ G $||\hat{z}$ where the electron spin states show an anticrossing in the excited state (see circled region in Fig.~1a).  At the excited state level anticrossing (ESLAC), hyperfine interactions in the excited state of the NV center enable resonant electron-nuclear spin flip-flops, effectively transferring the electron spin polarization to the $^{14}$N nuclear spin in a two-step process~\cite{Smeltzer09, Steiner10}.  As a result, in the absence of RF excitation, only a single line would be visible in the ESR spectrum at 1.45 GHz.  While working at the ESLAC simplifies the spectrum by eliminating multiple hyperfine lines, it introduces other complications.  Because low-frequency electron spin transitions are possible in the excited state, we avoid applying RF fields during optical illumination; instead, MW and RF excitation is pulsed prior to fluorescence detection (see pulse sequence in Fig.~2a).   The resulting data set (Fig.~2a) exhibits the same qualitative features as the low-field data, and the triangular pattern of ``missing resonances" is more pronounced.  

 
Multifrequency excitation is commonly employed to observe double resonance phenomena~\cite{He93b, Manson90}, and we note in passing that nuclear magnetic resonance (NMR) and electron-nuclear double resonance (ENDOR) features can arise in this system when $\omega_{RF}$ is close to the ground state $^{14}$N nuclear spin transition  $m_I = +1 \rightarrow m_I = 0$ (see Fig.~2b).   At the ESLAC, nuclear spin flips directly correlate to reduced fluorescence, creating the horizontal feature observed in Fig.~2b at the NMR  frequency $\omega_{RF} \approx (2\pi) 5.1$ MHz~\cite{Smeltzer09, Steiner10}.  This feature is particularly dark because the $10\mu$s pulse is close to a $\pi$ pulse on the nuclear spin transition, whereas other (time-averaged) transitions only depopulate the ground state by 50\%.  Moreover, for $\omega_{RF}$ near the NMR transition, the $m_I = 0$ state enhances Raman transitions between the $m_s = 0, m_I = +1 \rightarrow m_s = -1, m_I = 0$ states (see Fig.~2b inset).  Such transitions occur, for example, when $\omega_{MW} = \Delta - \omega_{RF} - A$, where $\Delta$ is the bare ESR frequency and $A = 2.2$ MHz is the $^{14}$N hyperfine splitting~\cite{Felton09}.  Indeed, the data exhibits weak ENDOR features in the vicinity of $\omega_{RF} \sim 5$ MHz that are roughly $2.2$ MHz lower in frequency than each multiphoton resonance (see Fig.~2b).

NMR and ENDOR resonances in the NV center have been previously examined at the ground state level-anti-crossing~\cite{He93b, Manson90}, and they play only a minor role in the overall features observed in the multifrequency data.  In particular, the three features we consider -- the low-frequency splitting, multiphoton resonances, and ``missing resonances" are not associated with nuclear spin transitions.  These features of our data involve only the electronic spin transitions, and thus result from dynamics of an effective two-level system.

\subsection{Numerical simulations}

To elucidate the origins of the features in the RF/MW spectra, we numerically simulate the dynamics of a spin subject to multifrequency excitation in a parameter regime similar to our experiments.  Specifically, we consider the evolution of the two-level system formed by the $m_s = 0$ and $m_s = -1$ spin sublevels governed by the Hamiltonian 
\begin{equation}
\hat{H} = \frac{\Delta}{2} \sigma_z + \mathbf{\sigma}  \cdot \left(\mathbf{\Omega_{MW}}\cos{(\omega_{MW}t)} + \mathbf{\Omega_{RF}}\cos{(\omega_{RF}t + \phi)}\right)
\label{hamiltonian}
\end{equation}
where $\hbar = 1$, $\mathbf{\sigma} = \{\sigma_x, \sigma_y, \sigma_z\}$ represents the usual Pauli matrices for the pseudo-spin-1/2 system 
and $\mathbf{\Omega_{MW}}$ and $\mathbf{\Omega_{RF}}$ are vectors parallel to the magnetic field orientation for the MW and RF excitation respectively.  Here, factors of $1/2$ are chosen so that the transverse orientations of $\Omega_{MW}$ and $\Omega_{RF}$ correspond to the angular frequency for resonant Rabi nutations.

Experimentally, the MW and RF magnetic fields are oriented at an angle to the NV axis, but our numerical simulations indicate that only one component of each vector is responsible for the observed resonances.  Fig.~3 shows calculations of the time-averaged population in $m_s = 0$ for a spin initially polarized into $m_s = 0$ and evolving according to Eq.~\ref{hamiltonian} for three different orientations of $\mathbf{\Omega_{MW}}$ and $\mathbf{\Omega_{RF}}$.  In particular, the features we observe in our experiments arise solely from the $\hat{x}$ component of $\mathbf{\Omega_{MW}}$ and the $\hat{z}$ component of $\mathbf{\Omega_{RF}}$.  
 
\begin{figure}[htbp] 
  \centering
  \includegraphics[width=3.5 in,height=4in,keepaspectratio]{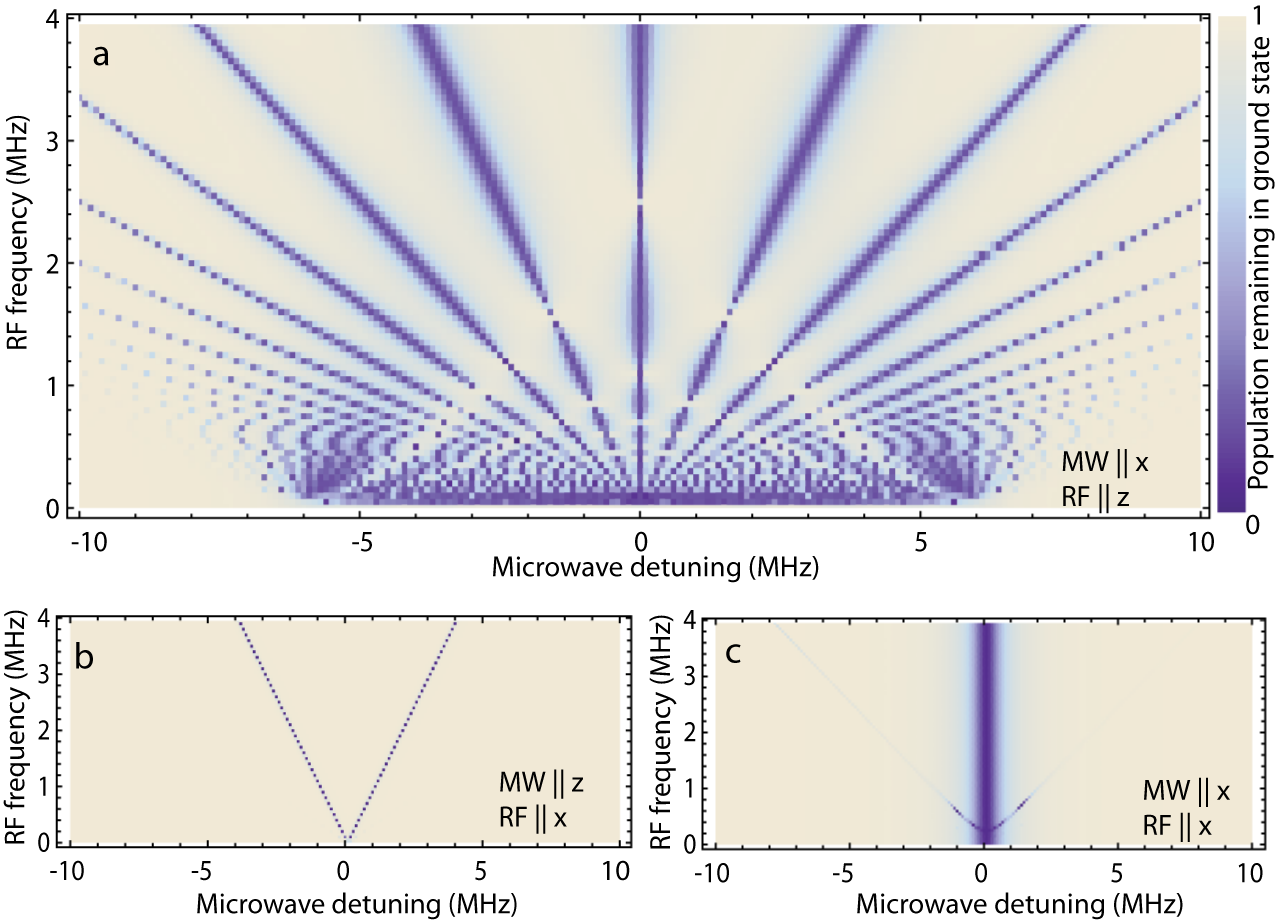}
  \label{numerical}
\caption{Numerical simulation of continous-wave multifrequency spectroscopy.  All simulations are performed using a fourth-order Runge-Kutta integration with a time step of 0.1 ns over a total time of 100$\mu$s.~\cite{ftnt} 
 In each case, $\Omega_{MW} = (2\pi) 0.5$ MHz, $\Omega_{RF} = (2\pi) 3$ MHz, and $\Delta = (2\pi) 100$ MHz, while the simulations average over time and $\phi$.   The microwave frequency is indicated as a detuning from the resonant frequency of $100$ MHz; this resonance frequency is far lower than the experimental resonance at $\sim$ few GHz, but the simulations are not sensitive to this parameter because the RWA is valid for the MW excitation.  The average population in $m_s = 0$ (for a spin initially polarized into $m_s = 0$) is shown in color scale, with darker color indicating lower population.   (a) $\mathbf{\Omega_{MW}} || \hat{x}$ and $\mathbf{\Omega_{RF}} || \hat{z}$. (b) $\mathbf{\Omega_{MW}} || \hat{z}$ and $\mathbf{\Omega_{RF}} || \hat{x}$. (c) $\mathbf{\Omega_{MW}} || \hat{x}$ and $\mathbf{\Omega_{RF}} || \hat{x}$.  Note that no spin transitions occur if both oscillating magnetic fields are oriented along $\hat{z}$.
}
\end{figure}

The dependence on MW and RF orientation can be explained qualitatively by considering the processes required to flip the spin by absorption of one MW and one RF photon~\cite{note}.  Absorption of a  photon polarized transverse to the NV axis must be accompanied by a spin flip; longitudinal photon absorption can be thought of as shifting energy without changing the spin state.  When both MW and RF fields are transverse, two-photon processes cannot drive a spin transition; only odd-photon-number processes are allowed, and they are strongly suppressed by the large detuning of their intermediate states.  As a result, $\Omega_{MW}||\hat{x}$, $\Omega_{RF}||\hat{x}$ only significantly drives the single-photon transition -- a weak three-photon process is also visible in Fig.~3c.  When allowed by selection rules (i.e. for $\Omega_{MW}||\hat{x}, \Omega_{RF}||\hat{z}$  or $\Omega_{MW}||\hat{z}, \Omega_{RF}||\hat{x}$) the two-photon transition rate will roughly scale as $\Omega_{MW}\Omega_{RF}/D$ where $D$ is the detuning of the intermediate state.   For $\Omega_{MW}||\hat{z}, \Omega_{RF}||\hat{x}$, $D\sim\omega_{MW}$, and multiphoton transitions are strongly suppressed (see Fig.~3b).  Only for $\Omega_{MW}||\hat{x}, \Omega_{RF}||\hat{z}$ is the detuning $D\approx\omega_{RF}$ small enough that strong two-photon (and higher) transitions occur (see  Fig.~3a).  Qualitative considerations thus substantiate the numerical simulations, indicating that the transverse MW and longitudinal RF magnetic field components are responsible for the features in our experimental data. 

\subsection{Theory}

While numerical techniques can confirm the behaviors of the system and help eludicate their origin, an  analytic explanation provides physical intuition for the features in the spectrum and may facilitate development of applications.   In particular, we analyze the large splitting in the ESR lines at low RF frequency, the multiphoton transitions at high RF frequency, and lastly the ``missing resonances" that appear for moderate RF frequencies.  
By use of quasistatic approximations and two-level system analysis in the interaction picture, we obtain a straightforward explanation of these phenomena.  

The features we seek to understand arise from specific orientations of the MW and RF field, so we begin by restricting Eq.~\ref{hamiltonian} to MW excitation perpendicular to the spin axis (transverse) and RF excitation along the spin axis (longitudinal).  Further simplification can be made by applying the RWA to the microwave excitation.  
Choosing a coordinate system to eliminate $\phi_{MW}$ and moving to a frame rotating with the microwave frequency yields
\begin{equation}
\hat{H} \approx \frac{\delta}{2}\sigma_z + \frac{\Omega_{MW}}{2}\sigma_x + \Omega_{RF}\cos{(\omega_{RF}t + \phi)}\sigma_z,
\label{RWA}
\end{equation}
where $\delta = \Delta - \omega_{MW}$.
The corrections (of order $\left(\frac{\Omega_{MW}}{\omega_{MW}}\right)^2$) are negligible for the parameter regimes we consider, where $\omega_{MW}\gg \Omega_{RF}\sim \omega_{RF} > \Omega_{MW}$.  For example, no difference is observed for numerical simulations that directly integrate Eq.~\ref{hamiltonian} and those use Eq.~\ref{RWA} for the parameters given in Fig.~3a.  Moreover, this Hamiltonian is now of a form that has recently received some attention because of its applications to superconducting qubits~\cite{Ashhab07, Son09}: the off-diagonal couplings are constant and the diagonal terms are modulated in time. Indeed, many of our observations are similar to dynamics of microwave driven superconducting devices~\cite{Nakamura01}.

The behavior at very low RF frequencies in Fig.~1c and Fig.~2a is readily obtained from Eq.~\ref{RWA} via a quasi-static picture of the RF field.  In our experiments, the spin population evolves coherently during a time $t_{max}\sim$ few $\mu$s set by the smaller of the dephasing time $T_2^*$ of the electron spin and the duration of the excitation pulses.  For  $\omega_{RF}\ll 1/t_{max}$, each experiment occurs under conditions where the energy splitting $\delta + 2 \Omega_{RF}\cos{(\omega_{RF}t +\phi)}$ is approximately constant with a random value ranging between $\delta -2 \Omega_{RF}$ and $\delta + 2 \Omega_{RF}$. Owing to the shape of the cosine, the extremal values are most probable, and the observed ESR spectrum appears split by $2\Omega_{RF}$.  Specifically, the probability of a given detuning $x$ is proportional to $1/\sqrt{4 \Omega_{RF}^2 - (x-\delta)^2}$, which, when convolved with a Lorentzian, is in good qualitative agreement with the line shape (see Fig.~4a).  Fig.~4b confirms that the observed splitting scales with the Rabi frequency $\Omega_{RF}$: solid lines illustrate the expected functional dependence $\Omega_{RF} \propto \sqrt{10^{\mathrm{dB}/10}}$.  This quasi-static splitting provides a convenient mechanism for characterizing the $\hat{z}$ component of the RF field, and enables the estimates of  $\Omega_{RF}$ we provide for our data.  Note that these quasi-static effects do not show up in the numerical simulation because dephasing is not included and the timescale for integration far exceeds 1/$\omega_{RF}$ for all calculated values of the RF frequency.  

\begin{figure}[bhtp] 
  \centering
  \includegraphics[width=3.5 in,height=6in,keepaspectratio]{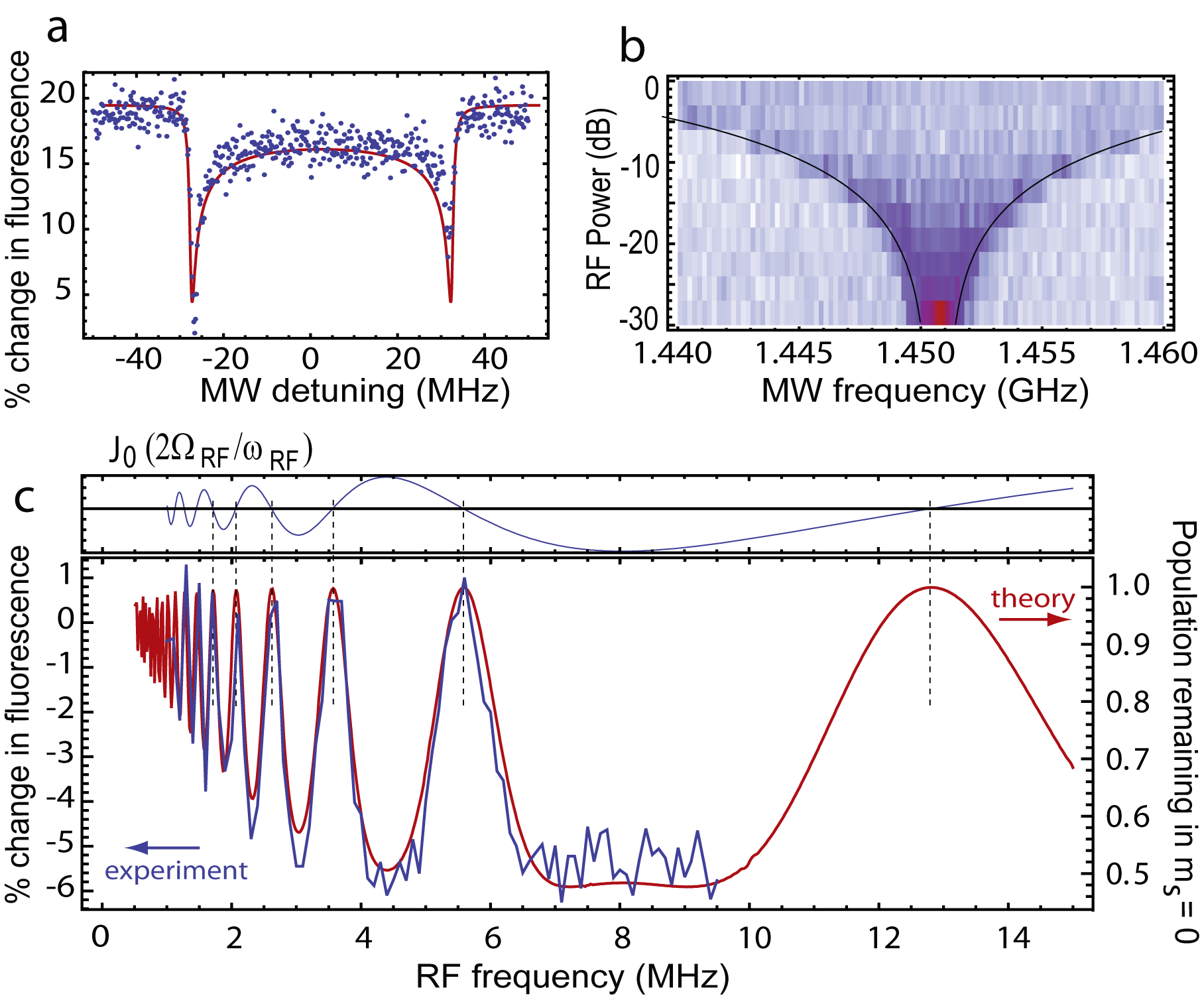}
  \label{analytic}
\caption{  (a) A single trace from Fig. 2a at $\omega_{RF}= (2\pi) 300$kHz, shown with the expected quasistatic lineshape (red line) for $\Omega_{RF} = (2\pi) 15 $ MHz convolved with a Lorenzian of HWHM 0.5 MHz. (b) MW spectrum vs RF power.  Solid lines illustrate functional dependence of Rabi frequency $\Omega_{RF}$ on RF power. (c) Resonant MW excitation ($\omega_{MW} \approx \Delta$) vs RF frequency, illustrating CDT.  Experimental data obtained by averaging over the central 1 MHz line in data from Fig.~2a.  Theory shows population in $m_s = 0$ after 5$\mu$s numerical integration of the Bloch equations for Eq.~\ref{RWA} using $\Omega_{MW} = (2\pi) 0.43$ MHz, $\Omega_{RF} = (2\pi)15.4$ MHz incorporating 0.5 MHz pure dephasing.  Upper curve indicates corresponding zeros in the Bessel function $J_0(2\Omega_{RF}/\omega{RF})$.
}
\end{figure}

Multiphoton transitions can be readily explained via Floquet theory~\cite{Shirley65}, but the parameter regime we investigate invites an alternate approach~\cite{Ashhab07, Son09} for analyzing the multiphoton and missing resonances.  These features occur when $\Omega_{MW}$ is the smallest parameter in our system, suggesting a natural division of the Hamiltonian into $\hat{H}=\hat{H}_{z}+\hat{H}_{MW}$, where $\hat{H}_{MW}$ is a weak perturbation.  The evolution arising from the diagonal components $\hat{H_z}$ can be found exactly, allowing us to move to the interaction picture, where the  Hamiltonian becomes
\begin{equation}
\tilde{H}_{MW} =\left(
 \begin{array}{clcr}
0&\sum_n A_n e^{-i(n\omega_{RF}-\delta)t}\\
\sum_n A_n e^{i(n\omega_{RF}-\delta)t}&0\\
\end{array}\right),
\label{int}
\end{equation}  
with $A_n = \frac{\Omega_{MW}}{2} J_n\left(\frac{2\Omega_{RF}}{\omega_{RF}}\right)$ (see e.g.~\cite{Ashhab07} for a full derivation).  In this picture, the basis states of the interaction Hamiltonian are spin eigenstates with time-varying phases $\propto -i\int \hat{H}_z dt$; transitions induced in the interaction picture thus correspond to spin flips in the Schrodinger picture.  

The interaction picture Hamiltonian (Eq.~\ref{int}) can be further simplified by again applying the rotating wave approximation. For weak MW excitation, the off-diagonal components of Eq.~\ref{int} are much smaller than $\omega_{RF}$, so only the term in the sum with $\delta - n \omega_{RF}\approx 0$ will lead to significant transition probability (the others will average to zero over the timescale of system response).  The different resonances $\delta = n\omega_{RF}$ thus correspond to the multiphoton transitions we observe in our data.  Furthermore, in the RWA each n-RF photon transition is driven with an effective Rabi frequency of $\Omega_{MW}J_n\left(\frac{2\Omega_{RF}}{\omega_{RF}}\right)$ when the resonance condition $\delta = n\omega_{RF}$ is satisfied.    The n-photon transition disappears for excitation frequencies $\omega_{RF}$ given by the roots of the n$^{th}$ order Bessel function, yielding the locations of the missing resonances. 



These missing resonances are related to so-called coherent destruction of tunneling (CDT), which is a strong-field phenomenon that only occurs when the Rabi frequency of an excitation exceeds its carrier frequency.  CDT arises in systems obeying a Hamiltonian of the form of Eq.~\ref{RWA} when the off-diagonal transition (in this case $\Omega_{MW}$) is small and the amplitude of the longitudinal oscillations ($\Omega_{RF}$) is larger than the oscillation frequency ($\omega_{RF}$).  Physically, CDT can be interpreted as a resonance in the Landau-Zener crossing between instantaneous eigenstates of Eq.~\ref{RWA} induced by a large amplitude oscillating detuning: it occurs when the amplitude and frequency of the oscillation conspire to precisely drive the system across the energy gap between instantaneous eigenstates during each half cycle.  CDT was first theoretically recognized for a particle confined to a double well~\cite{Grossman91}; it has since been observed in systems including cold atoms~\cite{Kierig08, Lignier07} and optical waveguides~\cite{ DellaValle07}.
 While recently there has been a theoretical description of CDT in NV centers~\cite{Wubs10}, the phenomenon has not, to our knowledge, been previously observed in this context.  

The spectra seen in our experiments closely match the predicted form for CDT resonances.  Fig.~4c shows the central ESR  transition ($\delta = 0$) from the data of Fig.~2a, as compared to the functional form of the expected Rabi frequency $\Omega_{MW}J_0(2\Omega_{RF}/\omega_{RF})$ and numerical integration of the Bloch equations for Eq.~\ref{RWA}.  The maxima in fluorescence (indicating suppression of ESR transitions) coincide with the Bessel function zeros for $\Omega_{RF} = (2\pi) 15.4$ MHz, and the shape of the curve fits closely with numerical simulations.  Note that the discrepancy between values for $\Omega_{RF}$ in (a) and (c) is required to fit the data, and likely arises from nonidealities in RF amplifier (Minicircuits ZHL-32A) frequency response in our experiments.  For higher-order transitions $\delta = n\omega$, the missing resonances coincide with zeros of $J_{\delta/\omega}(2\Omega_{RF}/\omega_{RF})$ as shown in the inset to Fig.~2a.  The excellent agreement between the frequencies of the observed missing resonances and theory confirms that they indeed arise from coherent destruction of tunneling. 


\section{Consequences and applications}  

While observations of multifrequency excitation of the NV center present an opportunity to explore quantum optical phenomena, they may also have relevance to applications in quantum information science.  In particular, our results highlight potential pitfalls in multiplexed electron-nuclear manipulation and point to possible applications using oriented microwave magnetic fields.  
Because the electron gyromagnetic ratio is several orders of magnitude larger than that of a nucleus, RF excitation intended for a nuclear spin will have a considerably larger effect on an  electronic spin.  Even with sequential pulses, care must be taken to control or account for phase accumulation from the longitudinal components of the RF magnetic field.  Similarly, attempts to simultaneously manipulate electron and nuclear spins must account for CDT suppression of electron spin transition rates.  

Most spin resonance experiments rely on transverse excitation, but longitudinal coupling offers an additional technique for spin manipulation.  In particular, it enables spin states to be brought quickly into and out of resonance without the need for fast magnetic field ramps.  The longitudinal selection rules involve no spin flips, so this type of excitation could potentially be used to turn on and off a static transverse coupling.  

As an example relevant in the context of the NV center in diamond, the nonsecular terms in a contact hyperfine Hamiltonian $ A(S_+I_- + S_-I+)$ provide such a static transverse coupling between the electron-nuclear spin states $|0, \downarrow\rangle$ and $|-1, \uparrow\rangle$. (Here, $\uparrow,\downarrow$ encodes a nuclear spin $I = 1/2$ projection along the NV axis, while numbers refer to the NV electronic spin projection $m_s$.)  While a static mangetic field could be used to bring the initial and final states into resonance, leading to oscillations between them, a longitudinal magnetic field with strength $\Omega \cos{\omega t}$ oscillating  at the energy difference between these states will also lead to electron-nuclear spin flip-flop transitions at a rate $\sim A\Omega/\omega$. With appropriate resonator~\cite{Alegre07} or stripline~\cite{Fuchs09} design, longitudinal excitation could thereby be used to implement electron-nuclear swaps or extend optical nuclear spin polarization to arbitrary magnetic field strengths~\cite{Smeltzer09}.  Although only one oscillating field is involved, this type of configuration mimics the multifrequency excitation discussed in this paper in that the nonsecular hyperfine terms resemble detuned MW excitation (in the RWA) and the applied longitudinal excitation plays the role of the RF by enabling resonant transitions.  

\section{Conclusion}

Simultaneous application of weak MW and strong RF excitation in a two-level system leads to a rich set of phenomena beyond the usual rotating wave approximation.  Using multifrequency excitation of the NV center in diamond, we observe quasi-static splitting, multiphoton resonances and coherent destruction of tunneling in the electronic spin degree of freedom.  Notably, the condition for strong-field behavior does not require Rabi frequencies comparable to the energy separation between spin states, making this regime easy to access experimentally.  Numerical simulations confirm our understanding that this behavior arises from combined longitudinal and transverse excitation, with resulting selection rules allowing resonant two-photon transitions.  These results indicate the importance of taking the orientation of applied microwaves into account when designing experiments; moreover, they invite exploration of control techniques based on longitudinal excitation.  Ultimately, a greater understanding of multifrequency excitation may inform new techniques for manipulating spin systems.

\acknowledgements{The authors acknowledge help from Kyle Enman in developing Matlab code for Floquet theory based calculations. The authors thank Jonathan Hodges, Jero Maze, and Gurudev Dutt for productive conversations.  This work was supported by a grant from Research Corporation and by funds provided by Bates College.  J.M. acknowledges support from HHMI.}

\bibstyle{unsrt}

\bibliography{paper2}

\begin{thebibliography}{32}
\bibitem[{ftn()}]{ftnt}
\bibinfo{note}{While Floquet theory provides another route to numerical
  simulation, at low RF frequencies the convergence of transition probabilities
  as a function of Floquet Hamiltonian dimension was sufficiently slow to
  warrant a direct numerical integration approach.}
\bibitem[{not()}]{note}
\bibinfo{note}{We note that the use of the word photon is not intended to
  indicate that the quantum aspects of the radiation are important. The RF and
  MW fields are classical, and the number of photons involves serves to
  indicate how the transition rates scale as powers of the applied fields.}
\end{thebibliography}

\end{document}